\documentclass[manuscript,screen,nonacm]{acmart}
\usepackage[utf8]{inputenc}
\usepackage[english]{babel}
\usepackage{booktabs}
\usepackage{array}
\usepackage{float}
\usepackage[justification=centering, skip=4pt]{caption}


\usepackage{tikz}
\usetikzlibrary{positioning, arrows.meta, fit, calc}
\definecolor{figtint}{HTML}{DCE7F3}
\definecolor{figaccent}{HTML}{33628F}
\definecolor{figmuted}{HTML}{5C6672}
\usepackage{csquotes}
\MakeOuterQuote{"}

\title{Skill Roulette: Measuring Skill Invocation Rates Across LLM Coding Agents}
\author{Jin Rae Lee}
\affiliation{%
  \institution{Singapore Management University}
  \city{Singapore}
  \country{Singapore}
}

\author{Jai Lal Lulla}
\affiliation{%
  \institution{Singapore Management University}
  \city{Singapore}
  \country{Singapore}
}

\author{Sebastian Baltes}
\affiliation{%
  \institution{Heidelberg University}
  \city{Heidelberg}
  \country{Germany}
}

\author{Christoph Treude}
\affiliation{%
  \institution{Singapore Management University}
  \city{Singapore}
  \country{Singapore}
}

\begin{document}

\begin{abstract}
Agent skills package procedures for large language model (LLM) coding agents. An agent selects a skill from its name and description before reading the procedure. It may complete the task without invoking a relevant skill. We investigate how reliably coding agents invoke relevant skills and which factors affect that decision. We estimate invocation rates by repeating trials with identical inputs; invocation rate is the proportion of trials in which the agent invokes the relevant skill. We test 9 non-directional hypotheses about skill metadata, competing skills, contextual instructions, and task wording and composition. Eight concern invocation; the ninth tests whether making a skill available changes the occurrence of an action that the task does not request. We use tasks from 4 software engineering categories in a synthetic repository, running each of 4 coding-agent harnesses with 2 models. The planned 20,400 trials include 60 repetitions per Baseline category and configuration, 60 more per category in Claude Code with Sonnet 5.5 that admit each category's task and skill, and 10 per experimental condition. Comparisons are specified before collection and evaluated within each model and harness configuration using Boschloo's exact test.
\end{abstract}
\keywords{agent skills, LLM coding agents, skill invocation, invocation rate, empirical software engineering}

\ccsdesc[500]{Software and its engineering~Development frameworks and environments}

\maketitle

\section{Introduction}

\begin{figure}[H]
\centering
\resizebox{\linewidth}{!}{%
\begin{tikzpicture}[
  x=1mm, y=1mm,
  box/.style={draw, rectangle, minimum height=11mm, align=center,
              inner sep=2pt, font=\ttfamily\scriptsize},
  boxhi/.style={box, draw=figaccent, line width=1.2pt, fill=figtint},
  card/.style={draw, rectangle, line width=0.5pt},
  split/.style={draw, line width=1.2pt},
  cardtext/.style={font=\ttfamily\scriptsize, inner sep=0pt},
  pool/.style={draw=black!35},
  stub/.style={thick},
  arrow/.style={-Stealth, thick},
  arrowdim/.style={-Stealth, thick, dashed, draw=figmuted},
  edge/.style={font=\ttfamily\tiny, text=figmuted},
  actor/.style={font=\ttfamily\tiny, text=figmuted}
]

\draw[pool] (21,27.5) rectangle (73,52.5);
\node[actor] at (47,50.5) {SKILL POOL};

\fill[black!8] (55,43.1) rectangle (69,48.1);
\draw[card]    (25,43.1) rectangle (69,48.1);
\draw[split]   (55,43.1) -- (55,48.1);
\node[cardtext]                at (40,45.6) {name + description};
\node[cardtext, text=figmuted] at (62,45.6) {body};

\fill[black!8] (55,37.3) rectangle (69,42.3);
\draw[card]    (25,37.3) rectangle (69,42.3);
\draw[split]   (55,37.3) -- (55,42.3);
\node[cardtext]                at (40,39.8) {name + description};
\node[cardtext, text=figmuted] at (62,39.8) {body};

\fill[black!8] (55,31.5) rectangle (69,36.5);
\draw[card]    (25,31.5) rectangle (69,36.5);
\draw[split]   (55,31.5) -- (55,36.5);
\node[cardtext]                at (40,34.0) {name + description};
\node[cardtext, text=figmuted] at (62,34.0) {body};

\node[box,   minimum width=15mm] (task) at (7.5,0)   {task};
\node[box,   minimum width=32mm] (ctx)  at (38,0)
  {{\tiny\textcolor{figmuted}{HARNESS}}\\[1pt] context assembly};
\node[boxhi, minimum width=37mm] (dec)  at (79.5,0)
  {{\tiny\textcolor{figmuted}{AGENT}}\\[1pt] invocation decision};
\node[box,   minimum width=32mm] (body) at (121,0)   {skill body loads};
\node[box,   minimum width=29mm] (exec) at (158.5,0)
  {{\tiny\textcolor{figmuted}{AGENT}}\\[1pt] task execution};

\draw[stub] (25,45.6) -- (18,45.6);
\draw[stub] (25,39.8) -- (18,39.8);
\draw[stub] (25,34.0) -- (18,34.0);
\draw[arrow] (18,45.6) -- (18,12) -- (38,12) -- (ctx.north);
\node[edge, anchor=south west] at (38.8,12.3) {every skill's metadata};

\draw[arrowdim] (69,39.8) -- (76,39.8) -- (76,15) -- (121,15) -- (body.north);
\node[edge, anchor=west] at (77,30) {the matched body};

\draw[arrow] (task.east) -- (ctx.west);
\draw[arrow] (ctx.east)  -- (dec.west);
\draw[arrow] (dec.east)  -- (body.west) node[edge, midway, above] {match};
\draw[arrow] (body.east) -- (exec.west);

\draw[arrow] (dec.south) -- ++(0,-7.5) -| (exec.south);
\node[edge, anchor=south] at (119,-12.3) {no match};

\end{tikzpicture}%
}
\caption{\textit{Skill invocation: a body loads only after the agent commits to the skill from name and description alone. H1--H8 measure invocation; H9 measures action occurrence.}}
\Description{Block diagram of skill invocation: a skill pool provides metadata to a harness for context assembly; the agent decides whether to invoke a skill; if matched, the skill body loads and the agent executes the task.}
\label{fig:invocation-mechanism}
\end{figure}
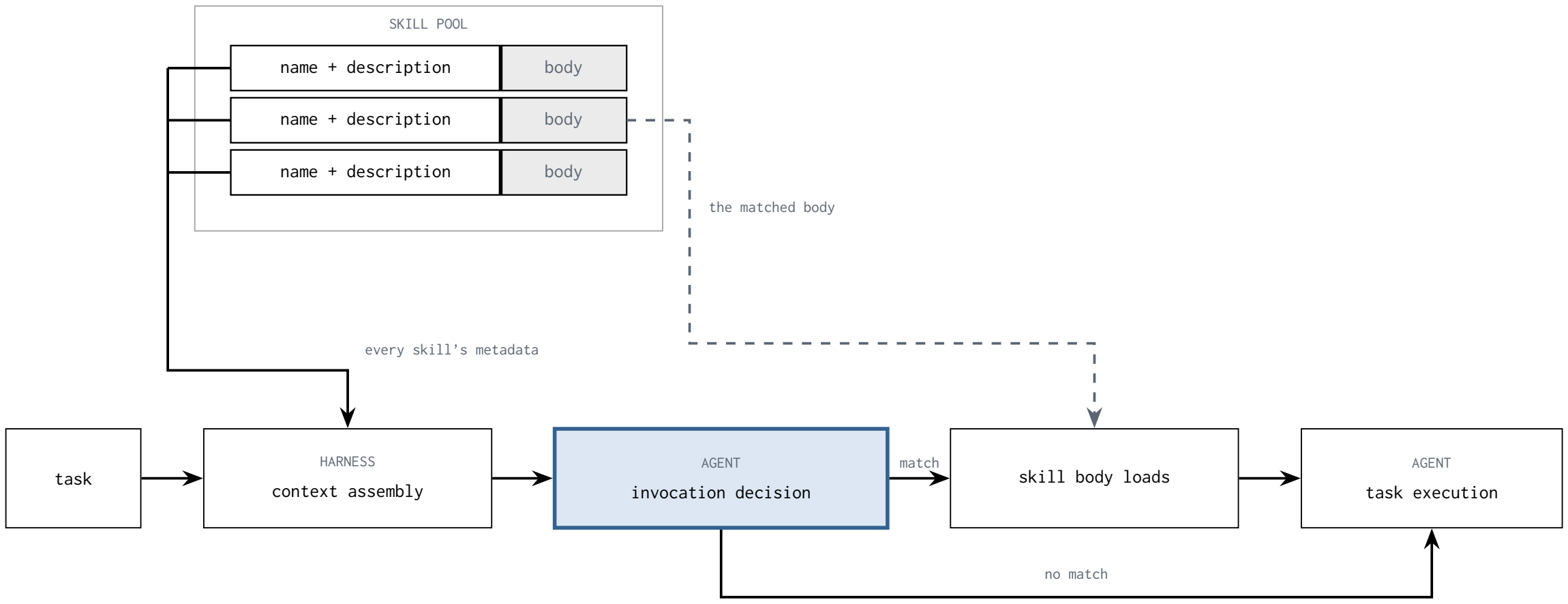

An agent skill is a file containing a procedure that a large language model (LLM) coding agent should follow for a class of tasks. The agent first reads the skill's name and description, assesses its relevance, and decides whether to load the body~\citep{anthropic2026agentskills}. The procedure can guide the agent only after the skill body loads (Figure~\ref{fig:invocation-mechanism}).

Successful task completion does not establish whether an agent invoked a relevant skill or followed its procedure. A trace can establish whether the skill was invoked, but deciding whether it should have been invoked requires an assessment of its relevance. When a missed invocation matters, a developer can supply the procedure directly or require the agent to invoke the skill. Understanding when such omissions occur is a prerequisite for detecting and correcting them in autonomous workflows.

Invocation reliability becomes particularly important when a workflow requires several procedures. If 4 invocation decisions are independent, each has a probability of 0.6, and there is no recovery, the probability that all 4 skills are invoked is about 13\%; at 0.5 each, it is about 6\%. Dependencies and recovery mechanisms can change these probabilities.

This study addresses whether an available, relevant skill is invoked reliably and which changes to its metadata, task, or context affect that decision. We estimate invocation rates by repeating trials with identical inputs.%

The study tests changes to skill metadata and pool composition (H1--H4), instructions and task wording (H5--H7), and the combination of tasks in a single prompt (H8). H9 asks whether making a skill available changes the occurrence of an unrequested action. We run tasks from 4 software engineering categories under 8 combinations of coding-agent harness and model. Each experimental condition has 10 repeated trials ($k=10$); the Baseline has 60 per category and configuration to check invocation for a clearly relevant skill.

We specify the task and skill construction, experimental conditions, and comparisons needed to measure invocation separately from task completion. We will release the repository, skill inventory, and trial records for replication with other harnesses and models.

\section{Background and related work}

\subsection{Skill configuration and discovery}

Skills configure how an agent should use its capabilities on a task. Training and fine-tuning develop the model's capabilities, while tools provide the actions available to it. A skill supplies a procedure that takes effect only if the agent invokes it and follows its instructions. Skills were one of eight configuration mechanisms documented across five agentic coding tools, and one of two supported by all five~\citep{galster2026configuring}. A study with a sampling frame of 40,585 repositories identified 2,430 skill artifacts in 547 repositories~\citep{galster2026dataset}. Marketplace inventories can be much larger; one grew from 2,179 to 40,285 listed skills in 20 days~\citep{ling2026agentskills}.
Skills are concentrated in a few of the 18 knowledge areas in the Software Engineering Body of Knowledge (SWEBOK)~\citep{washizaki2024swebok}. \citet{gao2026registryrepository} found that Software Construction accounted for 28.3\% of centralized skills and 22.1\% of personal-use skills, the largest share in both populations. Construction, Operations, Testing, and Configuration Management together accounted for over 40\% of centralized skills, with smaller shares in the remaining 14 areas. Metadata quality and naming overlap motivate our experiments: 26.4\% of skills sampled from public GitHub repositories had no routing description~\citep{gao2026skillreducer}, and 46.3\% of listings in one marketplace shared a normalized name with another listing~\citep{ling2026agentskills}. Shared names do not establish that procedures overlap or that near-duplicates occur together in a repository.

\citet{gao2026skillreducer} tested whether descriptions could be compressed without changing skill selection. Of 600 skills, 252 (42.0\%) preserved routing directly; restoring selected text or reverting to the original description preserved routing for all 536 eligible skills. The other 64 were excluded because their original descriptions did not trigger invocation~\citep{gao2026skillreducer}. A separate body evaluation compared original skills, compressed skills with on-demand references, and no skills on 5 generated tasks per skill. Mean scores across 600 skills improved by 2.8\% relative to the originals. These tests assessed preservation under compression and task performance, but did not report repeated identical-condition trials of invocation. We repeat each condition with the same task and pool to measure variation in invocation itself.

On SkillRouter's benchmark of 80,000 skills, hiding skill bodies reduced routing accuracy by 37--44 percentage points across dense retrieval and reranking baselines~\citep{zheng2026skillrouter}. SkillRouter retrieves and reranks candidates before presenting them to the agent, which must then decide whether to load a body from the available metadata. A coding-agent harness also controls which metadata enters the agent's context. We vary the harness to examine invocation under these different context-assembly mechanisms.

\subsection{Repeated agent behavior}

Repeated tool-use decisions provide a comparison for skill invocation. \citet{yagubyan2026consistent} studied 6 models on 19 tasks with deterministic simulated tools. Tool-sequence consistency was associated with task correctness, while argument consistency was not significantly associated with correctness. Skill invocation also depends on a selection decision, but then exposes a procedure that may affect later decisions. The tool-use findings do not establish how repeatable skill invocation is or whether agents correct an initial routing error after reading a skill body. \citet{yang2026ambiguity} varied task wording on a fixed task set and measured changes in function-level code generation. Their design measures repeated code-generation outcomes; ours measures whether a skill is invoked.

Repeated evaluations also distinguish occasional success from reliable task completion. In CAR-bench's hardest task category, frontier agents succeeded in all 3 repeated trials less than 50\% of the time, despite higher rates of success in at least 1 trial~\citep{kirmayr2026carbench}. \citet{raj2026consistency} proposed statistical methods for evaluating agent consistency and showed that trajectory measures detected failures that pass@1 obscured. We apply repeated trials to skill invocation.

\section{Methodology}

We vary the information available to an agent before it decides whether to invoke a skill. Each trial presents a task and a skill pool in a fresh copy of a synthetic repository. The design crosses 4 harnesses with 2 models and tests 9 hypotheses about metadata, instructions, and task prompts. H1--H8 measure invocation; H9 measures whether the agent performs an unrequested action. Table~\ref{tab:hypotheses-overview} summarizes the conditions and trial counts.

\subsection{Definitions}

\textbf{SWEBOK:} the Software Engineering Body of Knowledge, an IEEE Computer Society reference dividing the discipline into 18 knowledge areas~\citep{washizaki2024swebok}. This study draws its 4 task categories from 4 of those areas: Construction, Testing, Quality, and Security. Software Engineering Operations and Software Configuration Management, despite ranking among the areas with the largest empirical skill share after Construction~\citep{gao2026registryrepository}, are excluded: Both require artifacts (a deployment target, a monitoring system, a build pipeline) that this study's synthetic repository does not have.

\textbf{Target skill:} the skill manipulated by a hypothesis and evaluated for invocation. We author one per SWEBOK category, matched to that category's task.

\textbf{Skill pool:} the skills we place in the repository for a trial. By default, the pool contains 1 target skill and 3 fillers.

\textbf{Filler:} a skill in the trial's pool other than its target skill.

\subsection{Repository, tasks, and skills}

We author the tasks against the JavaScript HTTP server in the staged-project protocol of \citet{lulla2026buildvsbuy}. The server gains features in stages, and each stage starts from a partial implementation scaffolded to the point where that stage's feature must be added, with no residual package installations. We select the second stage because the initial server may be too simple for meaningful invocation tests, while later stages would make individual trials harder to interpret. We use the same snapshot throughout and do not repeat the trials in Python because programming language is not an experimental factor.%

\subsubsection{Authoring a task per category}

Each category's task asks the agent to perform an action from that category on the same repository snapshot. We reset the repository before each trial so that tasks do not inherit changes from earlier trials.

Each task must map to a single category without an equally plausible interpretation under another. Tasks must also have comparable scope to reduce differences in routing difficulty caused by task complexity. We manually assess the task prompts and compare mean output-token counts across categories within each configuration, without setting a threshold for the difference. Each task identifies an action category and the specific action required. A language model drafts the tasks against these criteria, and the authors review and revise them.

\subsubsection{Authoring skills}

A language model drafts skills using examples sampled from skills.sh~\citep{vercel2026skillssh} to guide their wording. The authors check each draft's form, information content, and length. We estimate length with an open-source approximation of the harness tokenizers and accept drafts only within the interquartile range of token counts in a representative sample of skills.sh listings. We release the accepted names, descriptions, and task texts with the analysis code.

Each category has 1 target skill with a name, a description, and a body containing generic instructions. The body does not specify a task procedure because the primary outcome is invocation. In the default pool, the other 3 categories' target skills serve as fillers. The Baseline checks whether the agent reliably invokes the matching target in this pool before the hypothesis trials begin.%

Unless a hypothesis specifies otherwise, the target is the only relevant skill in the pool, its name and description clearly match the task, and the project instruction file is empty. No additional instruction asks the agent to use skills.

\subsection{Handling filler invocations}

When a filler is invoked, we inspect the trial's response text and any reasoning summaries, check for accidental keyword overlap between the filler metadata and task, and assess whether the target remains relevant under the manipulation. We report this assessment with the filler invocation rate and retain the trial in every assigned comparison.

We treat accidental metadata overlap or a loss of target relevance as a possible problem with the experimental setup. Otherwise, the filler invocation is behavior observed under the condition. We retain both kinds of trial because excluding filler invocations could bias the comparison when a manipulation also reduces target invocation.

This treatment of filler invocations is the default for every hypothesis. Each subsection identifies any exception for a pool without fillers or an outcome that includes filler invocation.

We report invocations of skills that a harness ships with for each condition, separately from target and filler invocations.

\subsection{Models and harnesses}

Every trial runs non-interactively through a harness's software development kit (SDK) or command-line interface (CLI). We use 4 harnesses: Anthropic's Claude Code, OpenAI's Codex, Anomaly's OpenCode, and Mario Zechner's Pi. The models are Sonnet 5.5 and GPT-6.1 Sol. In every harness, each model runs at the default effort of its vendor's harness on October 10, 2026: medium for Sonnet 5.5 in Claude Code 2.1.296~\citep{anthropic2026claudecodemodel} and low for GPT-6.1 Sol in Codex 0.162.1~\citep{openai2026codexcatalog}. We configure the effort level through each harness's native option and record the setting in effect for every trial using the harness's own terminology. Vendors' effort labels need not represent equivalent settings.%

A harness assembles the agent's context and supplies its system instructions and tools. Crossing the models and harnesses lets us compare each model in its vendor's harness, in the other vendor's harness, and in model-agnostic harnesses. These comparisons include the combined effects of each harness's features.

Claude Code and Codex are vendor-specific harnesses. In an early-2026 survey of 906 readers of The Pragmatic Engineer, \citet{orosz2026aitooling} found that Claude Code was the most-used tool, having overtaken GitHub Copilot within 8 months of release. Codex, absent from the previous survey, had reached 60\% of Cursor's usage. Each harness uses its vendor's model by default. Anthropic describes Claude Code's design in terms of giving agents the tools that programmers use~\citep{anthropic2025agentsdk}. Codex emphasizes sandboxed execution and approval requirements~\citep{openai2026codexloop}. The cross-vendor configurations test invocation when each harness runs the other vendor's model.

OpenCode and Pi are open-source, model-agnostic harnesses with different built-in capabilities. OpenCode documents subagents, skills, and Model Context Protocol (MCP) server support~\citep{opencode2026docs}. Pi supports on-demand skills and built-in MCP with Codemode, while leaving subagents, plan mode, permission prompts, task lists, and background shell execution to extensions or external workflows~\citep{pidev2026}. Their inclusion lets us compare invocation under these different sets of capabilities.

All 4 harnesses support the Agent Skills format~\citep{anthropic2026agentskills,opencode2026docs,pidev2026,openai2026codexskills}. Each loads skill names and descriptions into the agent's context and loads a skill's full \texttt{SKILL.md} only when the agent invokes it. We place the same \texttt{SKILL.md} contents in \texttt{.claude/skills/} for Claude Code and in \texttt{.agents/skills/} for Codex, OpenCode, and Pi, the project locations each harness documents. Each harness runs as its current release at the time of collection with its shipped defaults, including any skills it ships with, except for the specified manipulations. We record the skills each harness ships with at the start of collection and after each harness update.%

Table~\ref{tab:configs} shows the 8 model and harness configurations. Six connect directly to the model provider. Claude Code with the OpenAI model and Codex with the Anthropic model use OpenRouter to translate between the vendors' protocols. These harnesses do not natively support the other vendor's application programming interface (API). Claude Code connects to OpenRouter's endpoint compatible with the Anthropic Messages API, and Codex connects to its endpoint compatible with the Responses API.%

\begin{table}[H]
\centering
{\setlength{\tabcolsep}{4pt}\renewcommand{\arraystretch}{1.1}
\begin{tabular}{l c c c c}
\toprule
 & \multicolumn{4}{c}{\textbf{Harness}} \\
\textbf{Model} & Claude Code & Codex & OpenCode & Pi \\
\midrule
Anthropic model & $\bullet$ & $\circ$ & $\bullet$ & $\bullet$ \\
OpenAI model & $\circ$ & $\bullet$ & $\bullet$ & $\bullet$ \\
\bottomrule
\end{tabular}}
\caption{\textit{The 8 harness-by-model configurations. Filled: direct to the model provider. Open: routed through OpenRouter.}}
\label{tab:configs}
\end{table}

\subsection{Baseline}
The Baseline measures invocation for a clearly relevant target in each of the Construction, Testing, Quality, and Security categories~\citep{washizaki2024swebok}, with the other 3 category skills as fillers. The four-skill pool matches the median for repositories with 1,000--9,000 stars and more than 1 skill ($n=313$) in \citet{gao2026registryrepository}'s replication package~\citep{gao2026replication}. Target names identify the qualified subject and operation, while descriptions state the subject, operation, and when to use the skill. Table~\ref{tab:baseline-example} gives the Construction example. The Baseline adds no manipulation.

\begin{table}[H]
\centering
{\setlength{\tabcolsep}{4pt}\renewcommand{\arraystretch}{1.2}
\begin{tabular}{l p{0.7\linewidth}}
\toprule
\textbf{Task} & "The server currently discards everything after the request line, so it has no way to read a request's headers. Make it read the headers sent with each request."\\
\addlinespace
\textbf{Target skill name} & Request header reading\\
\textbf{Target skill description} & Handles HTTP request headers. Reads and stores each header sent with the request. Use when a server needs to access header values.\\
\bottomrule
\end{tabular}}
\caption{\textit{Example task and target skill, for the Construction category.}}
\label{tab:baseline-example}
\end{table}

The Baseline checks task and skill pairings before the hypothesis trials. Only Claude Code with the Anthropic model must pass the admission gate. We use this as the reference configuration because Anthropic introduced the skills mechanism and Claude Code provides its native implementation. We require a near-ceiling invocation rate in this configuration to reduce ambiguity between poor task-skill matching and an effect of the manipulation. Each target must be invoked reliably with the other 3 categories' skills present as fillers.%

The gate requires 0 misses across 60 trials. For a fixed sample, this gives an exact one-sided 95\% Clopper--Pearson lower bound above the target invocation probability of 0.95~\citep{clopper1934confidence}. The required sample size is $n = \ln(1-0.95)/\ln(0.95) \approx 58.4$; we use 60. \citet{yeon2025certification} used a similar bound for tool-selection satisfaction probability. If a category fails, we check for wording or setup problems before revising or replacing its task or skill. A revised task repeats the gate in its own category. A revised skill repeats the gate in all 4 categories, because each target skill is a filler in the other 3 pools. We retain and report the trials of every gate, failed or passing. Tasks and skills are fixed once all 4 categories pass. We then run 60 new Baseline trials per category in this configuration before any hypothesis trials begin, and later comparisons use these trials as the Baseline. Conditioning admission on 0 misses biases the observed Baseline invocation rate upward, which would raise each support criterion's chance probability above the values in Table~\ref{tab:power-sensitivity}. We run the Baseline in the other 7 configurations without an admission requirement.%

\subsection{Hypotheses}
Every hypothesis below is non-directional: No test predicts which condition produces a higher or lower rate, only whether the manipulated factor has an effect.

Each support criterion compares a hypothesis's conditions with a reference condition. Where a hypothesis crosses 2 factors, we compare the levels of each factor with the other factor held fixed and report these differences and the interaction descriptively, with 95\% confidence intervals.%

\begin{table}[H]
\centering
{\setlength{\tabcolsep}{4pt}\renewcommand{\arraystretch}{1.1}
\begin{tabular}{l l >{\raggedright\arraybackslash}p{0.34\linewidth} >{\raggedright\arraybackslash}p{0.24\linewidth} r}
\toprule
 & \textbf{Group} & \textbf{Tests} & \textbf{Factor(s) and levels} & \textbf{Trials} \\
\midrule
Baseline & & Each category's target skill is invoked reliably with no manipulation applied. & None & 2,160 \\
\midrule
H1 & Skill/pool & Metadata specificity and field affect invocation rate. & Name level $\times$ description level (4 $\times$ 4) & 5,120 \\
\addlinespace
H2 & Skill/pool & Skills with overlapping scopes in a pool affect invocation rate. & Competitor name/description pairing (3) & 960 \\
\addlinespace
H3 & Skill/pool & Conflicting task categories in a skill's name and description affect invocation rate. & Conflict pairing $\times$ task category (6 $\times$ 2) & 960 \\
\addlinespace
H4 & Skill/pool & The number of filler skills in the pool affects invocation rate. & Skill pool size (4: 1, 6, 14, 140) & 1,280 \\
\addlinespace
H5 & Environment & The type of conflicting instruction and its delivery channel affect invocation rate. & Signal type $\times$ delivery channel (3 $\times$ 5) & 4,800 \\
\addlinespace
H6 & Environment & Instruction pressure to use skills affects invocation rate. & Pressure strength $\times$ target availability (4 $\times$ 2) & 2,560 \\
\addlinespace
H7 & Environment & How a task is framed affects invocation rate. & Task framing (5) & 1,920 \\
\addlinespace
H8 & What's asked & Combining tasks in a single prompt affects invocation of each task's skill. & Task composition (2: solo, composed) & 320 \\
\addlinespace
H9 & What's asked & A skill's presence affects whether its action occurs on tasks that do not explicitly request or require it. & Skill presence $\times$ action (2 $\times$ 2) & 320 \\
\midrule
\textbf{Total} & & & & \textbf{20,400} \\
\bottomrule
\end{tabular}}
\caption{\textit{All planned trials in the study, by the Baseline and each of the 9 hypotheses, with their conditions.}}
\label{tab:hypotheses-overview}
\end{table}

For most hypotheses, the trial count is the number of conditions multiplied by 4 task categories, 8 configurations, and $k$ trials. H3's 12 conditions already include the category comparisons. H7 adds a reference condition to its 5 variants. H8 uses 4 task orderings and collects only composed trials, reusing the Baseline for solo trials. H9 has 4 conditions in Construction only. The Baseline uses 60 trials per category and configuration, and 60 admission trials per category in Claude Code with Sonnet 5.5.

\subsubsection{H1 (Skill Metadata)}
\textit{Metadata specificity and the field carrying that information affect skill invocation rate.}

A skill author must decide what to put in the name and description, because the agent selects a skill before reading its body. Editing the body alone cannot change that initial decision when the metadata and other inputs remain unchanged. H1 varies the detail in each field to test its effect on invocation. We also report descriptively whether a detailed description compensates for a less informative name, or the reverse.

Each field has 4 levels, with each successive level adding information. We cross the name and description levels to form the 16 conditions in Table~\ref{tab:h1-grid}.

The lowest description level (1) uses "None" because the platform rejects an empty field~\citep{anthropic2026agentskills}. The lowest name level (A) uses a shared opaque token.

Description level (2) names the subject: "Handles HTTP request headers." Level (3) appends the operation: "Reads and stores each header sent with the request." Level (4) appends when to use the skill: "Use when a server needs to access header values." Levels (3) and (4) retain the sentences from the preceding level. Level (4) follows Anthropic's guidance to state both what a skill does and when to use it~\citep{anthropic2026agentskills}.

Name level (B) identifies the subject, "Headers"; level (C) adds a qualifier, "Request headers"; and level (D) adds the operation, "Request header reading".

Pairing name level (D) with description level (4) reproduces the Baseline's target metadata. Lower levels remove information from one or both fields. Table~\ref{tab:h1-grid} identifies each condition by its name and description levels.

\begin{table}[H]
\centering
{\setlength{\tabcolsep}{4pt}\renewcommand{\arraystretch}{1.1}
\begin{tabular}{l c c c c}
\toprule
 & \multicolumn{4}{c}{\textbf{Description}} \\
\textbf{Name} & (1) None & (2) Subject & (3) + Operation & (4) + When-to-use \\
\midrule
(A) Opaque & A1 & A2 & A3 & A4 \\
(B) Subject & B1 & B2 & B3 & B4 \\
(C) + Qualifier & C1 & C2 & C3 & C4 \\
(D) + Operation & D1 & D2 & D3 & \textbf{D4} \\
\bottomrule
\end{tabular}}
\caption{\textit{H1 (Skill Metadata): name level (row, A--D) and description level (column, 1--4) identifying each pairing.}}
\label{tab:h1-grid}
\end{table}

H1's support criterion is met for a category and configuration if the invocation rates of at least 3 of the 15 cells other than D4 differ significantly from the Baseline. We compare name levels within each column of Table~\ref{tab:h1-grid}, where the description level is fixed, and description levels within each row, where the name level is fixed.

\begin{table}[H]
\centering
{\small
\setlength{\tabcolsep}{6pt}\renewcommand{\arraystretch}{1.15}
\begin{tabular}{@{}p{2.9cm}p{0.75\linewidth}@{}}
\toprule
\textbf{Factors} & Name level $\times$ description level\\
\textbf{Levels} & 4 $\times$ 4, defined in Table~\ref{tab:h1-grid}\\
\textbf{Controls} & Skill names do not include names of implementation techniques or mechanisms\\
\textbf{Support criterion} & Invocation rates of at least 3 of the 15 cells other than D4 differ significantly from the Baseline (Boschloo's exact test)\\
\bottomrule
\end{tabular}}
\end{table}

\subsubsection{H2 (Skill Competition)}
\textit{Skills with overlapping scopes in a pool affect invocation rate.}

H2 tests how competition from near-duplicate skills changes invocation. Each pool contains 4 entries describing the same underlying skill, with no unrelated fillers. Every entry corresponds to an H1 condition in which it was the only relevant skill among 3 fillers. Comparing its H1 and H2 invocation rates measures the effect of replacing those fillers with competitors.

All 3 pool configurations (P1--P3) use name levels (A)--(D), giving each entry in a pool a distinct name. They differ in how the description levels are paired with those names.

P1 uses the same level-(4) description for all 4 entries and varies the name from (A) to (D). Differences in invocation rates measure the contribution of the name when the description is fixed at the documented standard.

P2 increases name and description detail together, using A1, B2, C3, and D4. Each entry adds information to both fields.

P3 pairs more detailed names with less detailed descriptions, using A4, B3, C2, and D1. Table~\ref{tab:h2-grid} shows the entries in all 3 pool configurations.

\begin{table}[H]
\centering
{\setlength{\tabcolsep}{4pt}\renewcommand{\arraystretch}{1.1}
\begin{tabular}{l c c c c}
\toprule
 & \multicolumn{4}{c}{\textbf{Description}} \\
\textbf{Name} & (1) None & (2) Subject & (3) + Operation & (4) + When-to-use \\
\midrule
(A) Opaque & P2 & --- & --- & P1, P3 \\
(B) Subject & --- & P2 & P3 & P1 \\
(C) + Qualifier & --- & P3 & P2 & P1 \\
(D) + Operation & P3 & --- & --- & P1, P2 \\
\bottomrule
\end{tabular}}
\caption{\textit{H2 (Skill Competition): which pool configuration draws an entry from each (name, description) pair in H1's grid (Table~\ref{tab:h1-grid}, same rows and columns). A pair listing 2 pool configurations is reused in both.}}
\label{tab:h2-grid}
\end{table}

More than 1 entry can be invoked in an H2 trial. The inclusive invocation measure counts each entry that is invoked, regardless of other invocations or their order. We also record sole invocation, which counts an entry only when no other entry is invoked in the trial. We retain both measures because some workflows allow several skills to be used together, while others require exactly one.

For each skill entry, we compare its 10 H1 reference trials with its 10 H2 competition trials using Boschloo's exact test. There are 12 comparisons, some sharing H2 trials or H1 reference data. H2's support criterion is met for a category and configuration if the inclusive invocation rates of at least 3 of the 12 entries differ significantly between the 2 conditions.%

\begin{table}[H]
\centering
{\small
\setlength{\tabcolsep}{6pt}\renewcommand{\arraystretch}{1.15}
\begin{tabular}{@{}p{2.9cm}p{0.75\linewidth}@{}}
\toprule
\textbf{Factors} & Name/description pairing across near-duplicate competitors\\
\textbf{Levels} & 3 pool configurations (P1--P3), defined in Table~\ref{tab:h2-grid}\\
\textbf{Controls} & 4 near-duplicate entries per pool; no fillers\\
\textbf{Support criterion} & Inclusive invocation rates of at least 3 of the 12 entries differ significantly between the H1 reference and H2 competition conditions (Boschloo's exact test)\\
\bottomrule
\end{tabular}}
\end{table}

\subsubsection{H3 (Metadata Conflict)}
\textit{Conflicting task categories in a skill's name and description affect invocation rate.}

Independent revisions can leave a skill's name and description pointing to different tasks. H3 tests whether conflicting categories in these fields affect invocation.

We create conflicting metadata by taking a name from one task category and a description from another. Both fields retain the Baseline's level of detail: The name gives a qualified subject and operation, and the description gives a subject, operation, and when to use the skill. Only the category differs.

We test all 6 pairs of SWEBOK categories (Table~\ref{tab:h3-grid}). For each pair, one entry combines the first category's name with the second category's description; the other combines the second category's name with the first category's description. The pool also contains 2 fillers from the categories outside the pair, with names and descriptions that agree.

We run each category's task with this pool. For one task, the first entry matches by name and the second by description; for the other task, these roles reverse.

\begin{table}[H]
\centering
{\setlength{\tabcolsep}{4pt}\renewcommand{\arraystretch}{1.1}
\begin{tabular}{l c c c}
\toprule
 & \textbf{Testing} & \textbf{Quality} & \textbf{Security} \\
\midrule
\textbf{Construction} & 1 & 2 & 3 \\
\textbf{Testing} & & 4 & 5 \\
\textbf{Quality} & & & 6 \\
\bottomrule
\end{tabular}}
\caption{\textit{H3 (Metadata Conflict): the 6 conflict pairings. Each conflict pairing is run twice, once under each category's own task, using the same 2 entries both times.}}
\label{tab:h3-grid}
\end{table}

Both conflicting entries can be invoked in the same trial. We use H2's inclusive and sole-invocation measures and report whether the name-matching entry, the description-matching entry, both, or neither was invoked. We compare the 2 entries' rates descriptively to show which field better predicts selection when the fields disagree.

H3's support criterion is met for a category and configuration if the invocation rates of at least 2 of the 6 entries run under the category's task differ significantly from that category's Baseline.

\begin{table}[H]
\centering
{\small
\setlength{\tabcolsep}{6pt}\renewcommand{\arraystretch}{1.15}
\begin{tabular}{@{}p{2.9cm}p{0.75\linewidth}@{}}
\toprule
\textbf{Factors} & Conflict pairing; task\\
\textbf{Levels} & 6 conflict pairings, defined in Table~\ref{tab:h3-grid}, each run twice\\
\textbf{Controls} & Fillers fixed per conflict pairing and non-conflicting\\
\textbf{Support criterion} & Invocation rates of at least 2 of the 6 entries run under a category's task differ significantly from that category's Baseline (Boschloo's exact test)\\
\bottomrule
\end{tabular}}
\end{table}

\subsubsection{H4 (Skill Pool)}
\textit{The number of filler skills in the pool affects skill invocation rate.}

Adding skills to a repository may change how often existing skills are invoked. H4 tests this effect at 4 pool sizes.

Pool size includes the target and all fillers. We choose 3 levels from the skills-per-repository distribution in \citet{gao2026registryrepository}'s replication package~\citep{gao2026replication} and add a larger stress condition. The released dataset excludes repositories above the authors' outlier cutoff of 18 skills. Fillers added to the Baseline pool take their names and descriptions from skills.sh~\citep{vercel2026skillssh} listings whose scope does not overlap any of the 4 target skills. Their bodies contain generic instructions in the format of the target skills' bodies. These fillers differ in content from the Baseline's fillers, which are the other categories' target skills, but none competes with the target task. We select only listings within the token-length limits of the authored skills.%

Level 1 contains only the target, corresponding to the distribution's 25th percentile. Level 2 adds 2 fillers to the Baseline pool, for 6 skills at the 75th percentile. Level 3 adds 8 fillers to Level 2, for 14 skills, just above the filtered distribution's outlier threshold of 13.5. Level 4 adds 126 fillers to Level 3, for 140 skills, as an intentionally extreme stress condition, testing whether effects observed at smaller sizes continue or saturate.

H4's support criterion is met for a category and configuration if the invocation rates of at least 2 of the 4 levels differ significantly from the Baseline.

\begin{table}[H]
\centering
{\small
\setlength{\tabcolsep}{6pt}\renewcommand{\arraystretch}{1.15}
\begin{tabular}{@{}p{2.9cm}p{0.75\linewidth}@{}}
\toprule
\textbf{Factors} & Pool size\\
\textbf{Levels} & (1, 6, 14, 140), defined above\\
\textbf{Controls} & Fillers (non-competing); pool delivery location\\
\textbf{Support criterion} & Invocation rates of at least 2 of the 4 levels differ significantly from the Baseline (Boschloo's exact test)\\
\bottomrule
\end{tabular}}
\end{table}

\subsubsection{H5 (Signal and Source)}
\textit{The type of conflicting instruction and its delivery channel affect skill invocation rate.}

Project instructions can discourage skill use directly or contradict the requested task. We call the added instruction a signal and vary both its type and its delivery channel to test their effects on invocation in H5.

We use 3 signal types. Type (1) discourages skills and tools with the same instruction in every category: "Do not use skills or tools for this if you can do it yourself." Type (2) states a category-specific project convention that contradicts the requested task, without mentioning skills or tools. The Construction example is "Request handling reads only the request line and does not read or store headers." Type (3) concatenates the type-(2) instruction and the type-(1) instruction, in that order, without changing either sentence.%

The delivery channels are (A) a code comment in a file that the agent works on, (B) a project document that it is directed to read, such as \texttt{CONTRIBUTING.md}, (C) the project instruction file loaded by the harness, such as \texttt{AGENTS.md} or \texttt{CLAUDE.md}, (D) the task prompt, and (E) the system prompt.

Table~\ref{tab:h5-grid} crosses the signal types and channels. The Baseline provides the comparison without a restrictive signal. Channel (C) places the signal in the otherwise empty project instruction file.

\begin{table}[H]
\centering
{\setlength{\tabcolsep}{4pt}\renewcommand{\arraystretch}{1.1}
\begin{tabular}{l c c c c c}
\toprule
 & \multicolumn{5}{c}{\textbf{Channel}} \\
\textbf{Signal type} & (A) Comment & (B) Doc file & (C) Instruction file & (D) Prompt & (E) System prompt \\
\midrule
(1) Discouraging & A1 & B1 & C1 & D1 & E1 \\
(2) Conflicting & A2 & B2 & C2 & D2 & E2 \\
(3) Both & A3 & B3 & C3 & D3 & E3 \\
\bottomrule
\end{tabular}}
\caption{\textit{H5 (Signal and Source): the crossed conditions, e.g. E3 = both lines via the system prompt.}}
\label{tab:h5-grid}
\end{table}

Only channel (E) changes the system prompt. Claude Code and Pi add text to their system prompt with the \texttt{--append-system-prompt} flag~\citep{anthropic2026clireference, pi2026clidocs}, and OpenCode adds it through the \texttt{system} field of a session request~\citep{opencode2026requestsource}. Codex does not natively support appending to its system prompt and only allows its built-in instructions to be replaced~\citep{openai2026codexconfig}. For Codex, we copy the built-in instructions from its first model request for each model and version, append the signal, and supply the result as the replacement. Before collection, we compare each harness's first model request with and without the signal in every configuration and confirm that the signal is the only difference.%

H5's support criterion is met for a category and configuration if the invocation rates of at least 3 of the 15 conditions differ significantly from the Baseline. We compare signal types within each column of Table~\ref{tab:h5-grid}, where the channel is fixed, and channels within each row, where the signal type is fixed. In channels (A) and (B), the agent sees the signal only when it reads the file containing it, so we record whether that read precedes its first skill invocation.

\begin{table}[H]
\centering
{\small
\setlength{\tabcolsep}{6pt}\renewcommand{\arraystretch}{1.15}
\begin{tabular}{@{}p{2.9cm}p{0.75\linewidth}@{}}
\toprule
\textbf{Factors} & Signal type $\times$ delivery channel\\
\textbf{Levels} & 3 $\times$ 5\\
\textbf{Controls} & Signal wording, fixed per type across channels; filler skills\\
\textbf{Support criterion} & Invocation rates of at least 3 of the 15 conditions differ significantly from the Baseline (Boschloo's exact test)\\
\bottomrule
\end{tabular}}
\end{table}

\subsubsection{H6 (Instruction Pressure)}
\textit{Instruction pressure to use skills affects skill invocation rate.}

Instructions to use skills are a possible remedy for missed invocations, but may also prompt an agent to invoke an irrelevant skill. H6 tests how instruction strength changes target and filler invocation rates.

For each category and configuration, H6 uses the H1 entry with the observed invocation rate closest to 50\%. If several entries are equally close, we select the first in alphabetical order of its condition label in Table~\ref{tab:h1-grid}. This selection leaves room to observe either an increase or a decrease. We test it in a pool with 3 fillers and compare instruction levels by the target invocation rate. We also test the same instructions with only the 3 fillers present, measuring the proportion of trials in which any filler is invoked. The selected H1 entry may differ between configurations.

We apply 4 instruction levels in both target-availability conditions. Level 1 adds no instruction. Level 2 adds "Use skills where relevant." Level 3 adds "You must use available skills." Level 4 adds "You must use the \texttt{<skill-name>} skill." Levels 2 and 3 use generic instructions; level 4 tests explicit naming, even when the named target is absent.

H6's support criterion is met for a category and configuration if at least 2 of its 6 comparisons with level 1 are significant: the target invocation rates at levels 2--4 with the target present, and the filler invocation rates at levels 2--4 with the target absent.

We analyze the target-present and target-absent conditions separately. Within each configuration, we also compare reasoning-token counts between the 2 conditions as a measure of deliberation cost, grouping trials by whether any skill is invoked.

The target-absent condition is an exception to the default filler handling. Because no pool member fits the task, we measure filler invocation without applying the filler check. Trials in which more than 1 filler is invoked are reported as such.

\begin{table}[H]
\centering
{\small
\setlength{\tabcolsep}{6pt}\renewcommand{\arraystretch}{1.15}
\begin{tabular}{@{}p{2.9cm}p{0.75\linewidth}@{}}
\toprule
\textbf{Factors} & Instruction-pressure strength $\times$ target availability\\
\textbf{Levels} & Instruction levels 1--4 $\times$ (present, absent)\\
\textbf{Controls} & 3 fillers per pool\\
\textbf{Support criterion} & At least 2 of 6 comparisons with level 1 are significant: target invocation rates at levels 2--4 (target present) and filler invocation rates at levels 2--4 (target absent) (Boschloo's exact test)\\
\bottomrule
\end{tabular}}
\end{table}

\subsubsection{H7 (Task Framing)}
\textit{How a task is framed affects skill invocation rate.}

A skill author cannot control how a user phrases the task. H7 tests whether changes in task wording affect invocation when the skill metadata is fixed.

H7 uses 5 variants of the Baseline task prompt and the same H1 selection rule as H6: the entry with an observed invocation rate closest to 50\% in each category and configuration. A reference rate near 50\% leaves room to detect increases that a near-ceiling Baseline could obscure. As the reference, we collect $k$ new trials of the selected entry with the unmodified Baseline prompt.%

Each variant makes a single edit to the Baseline prompt and is compared with these reference trials. Condition (1) removes the action clause, leaving only the symptom, to test an underspecified request. Condition (2) names a technique to test whether specifying the implementation changes the invocation rate. Condition (3) adds urgency. Condition (4) asks the agent to gather information before acting. Condition (5) adds an unrelated action to test invocation when the prompt contains another request. Table~\ref{tab:h7-examples} gives the Construction examples.

\begin{table}[H]
\centering
{\setlength{\tabcolsep}{4pt}\renewcommand{\arraystretch}{1.1}
\begin{tabular}{c p{0.7\linewidth}}
\toprule
\textbf{\#} & \textbf{Example (Construction)} \\
\midrule
1 & Baseline minus the action clause: "The server currently discards everything after the request line, so it has no way to read a request's headers." \\
2 & Baseline plus: "...by parsing each line into a name and value, split on the first colon." \\
3 & Baseline plus: "Please fix this as soon as possible." \\
4 & Baseline plus: "Collect all the relevant information about this before acting." \\
5 & Baseline plus: "Also add a \texttt{.gitignore} entry for common editor swap files." \\
\bottomrule
\end{tabular}}
\caption{\textit{H7 (Task Framing): example task text for all 5 conditions, shown for the Construction category.}}
\label{tab:h7-examples}
\end{table}

H7's support criterion is met for a category and configuration if the invocation rates of at least 2 of the 5 variants differ significantly from the reference.

\begin{table}[H]
\centering
{\small
\setlength{\tabcolsep}{6pt}\renewcommand{\arraystretch}{1.15}
\begin{tabular}{@{}p{2.9cm}p{0.75\linewidth}@{}}
\toprule
\textbf{Factors} & Task wording/framing\\
\textbf{Levels} & 5 conditions, defined above\\
\textbf{Controls} & Skill name and description, identical across all conditions\\
\textbf{Support criterion} & Invocation rates of at least 2 of the 5 variants differ significantly from the reference (Boschloo's exact test)\\
\bottomrule
\end{tabular}}
\end{table}

\subsubsection{H8 (Task Composition)}
\textit{Combining tasks in a single prompt affects invocation of each task's skill.}

H8 tests whether an agent still invokes each relevant skill when a user combines several tasks in a single prompt. We concatenate the 4 category tasks and compare new trials for this composed condition with the Baseline's solo trials. The skill pool remains unchanged, but every entry now matches a requested task. We therefore count invocations toward each skill's rate and the joint pass rate without applying the filler check.

Task order within the composed prompt is balanced by a Latin square: Each task occupies each position exactly once across the orderings in Table~\ref{tab:h8-orderings}.

\begin{table}[H]
\centering
{\setlength{\tabcolsep}{4pt}\renewcommand{\arraystretch}{1.1}
\begin{tabular}{c l l l l}
\toprule
\textbf{Ordering} & \textbf{Position 1} & \textbf{Position 2} & \textbf{Position 3} & \textbf{Position 4} \\
\midrule
1 & Construction & Testing & Quality & Security \\
2 & Testing & Quality & Security & Construction \\
3 & Quality & Security & Construction & Testing \\
4 & Security & Construction & Testing & Quality \\
\bottomrule
\end{tabular}}
\caption{\textit{H8 (Task Composition): the 4 Latin-square task orderings.}}
\label{tab:h8-orderings}
\end{table}

We estimate each skill's invocation rate from all 40 composed trials per configuration: 10 repetitions of each ordering in Table~\ref{tab:h8-orderings}. Boschloo's exact test compares each rate with the same skill's Baseline solo rate. H8's support criterion is met for a configuration if the rates of at least 2 of the 4 skills differ significantly from their Baseline solo rates. We also report the proportion of composed trials in which all 4 skills are invoked as a descriptive joint pass rate. For each composed trial, we describe the order in which skills are invoked, including which skill is invoked first.

\begin{table}[H]
\centering
{\small
\setlength{\tabcolsep}{6pt}\renewcommand{\arraystretch}{1.15}
\begin{tabular}{@{}p{2.9cm}p{0.75\linewidth}@{}}
\toprule
\textbf{Factors} & Task composition\\
\textbf{Levels} & Solo (the Baseline's trials), composed (the Baseline's tasks concatenated, order balanced per Table~\ref{tab:h8-orderings})\\
\textbf{Outcome} & Each skill's own invocation rate under composition\\
\textbf{Controls} & Skill pool and task text (reused from the Baseline); skill name and description\\
\textbf{Support criterion} & Composition-condition rates of at least 2 of the 4 skills differ significantly from their Baseline solo rates (Boschloo's exact test)\\
\bottomrule
\end{tabular}}
\end{table}

\subsubsection{H9 (Task Scope)}
\textit{A skill's presence affects whether its action occurs on tasks that do not explicitly request or require it.}

Skill availability may affect agent behavior even when the task does not call for the skill: its metadata is visible before invocation, and its procedure may influence subsequent actions if invoked. H9 therefore tests whether making a skill available changes the occurrence of an action that the task does not request. We examine running the project's test suite and updating its docstrings after the unchanged Construction task (Table~\ref{tab:baseline-example}). Both actions are plausible after a code change, and testing both reduces reliance on a single action that may already occur in nearly every trial without a skill. For each action, we compare a pool containing only the relevant skill with an empty pool. For H9, we add a test script with tests for the existing routes and a docstring on one existing function in a file that the Construction task modifies. Both arms use this version of the repository. We use Construction because the other category tasks do not require implementing new functionality. Test execution is coded from the tool-call trace as a command that runs the test script. A docstring update is coded from the tool-call trace as an edit that adds or changes a docstring. Both actions are coded whether or not the skill was invoked.%

The comparison estimates the effect of making a skill available. Its name and description are visible before invocation, and loading its body may later affect the agent's actions. Counting actions regardless of invocation includes both possibilities.

H9's support criterion is met if action occurrence differs significantly between skill-present and skill-absent trials, with the two actions pooled into a single comparison of 20 trials per condition. This pooled comparison gives equal weight to the two actions; we also report the effect for each action descriptively. Each trial is scored only for the action it was run for. We also report invocation descriptively.

\begin{table}[H]
\centering
{\small
\setlength{\tabcolsep}{6pt}\renewcommand{\arraystretch}{1.15}
\begin{tabular}{@{}p{2.9cm}p{0.75\linewidth}@{}}
\toprule
\textbf{Factors} & Skill presence $\times$ action\\
\textbf{Levels} & (Absent, present) $\times$ (test execution, docstring update)\\
\textbf{Controls} & No fillers\\
\textbf{Support criterion} & Occurrence rate differs significantly between skill-absent and skill-present, both actions pooled (Boschloo's exact test)\\
\bottomrule
\end{tabular}}
\end{table}

\subsection{Analysis plan}
\label{sec:analysis}

\subsubsection{Statistical tests}

Every confirmatory outcome in the study is binary: whether any skill was invoked, whether the invoked skill was the target, whether more than 1 distinct skill was invoked, and whether an action occurred. Every comparison specified by a hypothesis's support criterion is tested separately within each of the 8 configurations of Table~\ref{tab:configs}, using that hypothesis's category grouping. H8 evaluates its 4 skills together within each configuration, and H9 uses Construction only.

Every comparison is between independent groups and uses Boschloo's exact test at $\alpha=0.05$. If both groups contain only successes or only failures, $p=1$. Boschloo's two-sided $p$-value doubles the smaller one-sided $p$-value. The 95\% confidence interval combines two one-sided 97.5\% bounds from the unconditional exact test ordered by Fisher's mid-$p$-value~\citep{fay2021twosample}.

The design favors breadth: it tests 9 hypotheses across 4 task categories and 8 configurations, with $k$ trials per condition, for 20,400 trials including the Baseline. Every comparison named in a support criterion has equal standing, and the study sets no minimum difference worth detecting. With $k=10$, two conditions differ significantly only when their rates differ by at least 0.4, and a condition compared with a Baseline of 60 invocations in 60 trials differs significantly at a rate of 0.8 or lower. We compute the chance probabilities in Table~\ref{tab:power-sensitivity} exactly from the comparisons each support criterion uses, evaluating comparisons that share reference trials jointly, and take the largest value over the common invocation rate.

To complement the confirmatory rate comparisons with qualitative evidence about agent behavior, for each hypothesis we select for exploratory review up to 2 comparisons named in its support criterion, taking those with the smallest $p$-values between 0.05 and 0.20 across all categories and configurations. Where comparisons tie, we select by which trials' traces show the most informative handling of the target skill. We review the tool-call traces, response text, and any reasoning summaries of their trials for how the agent handled the target skill. These reviews are reported separately from the support counts as exploratory results.%

\begin{table}[H]
\centering
{\small\setlength{\tabcolsep}{4pt}\renewcommand{\arraystretch}{1.1}
\begin{tabular}{l l c c c c c}
\toprule
\textbf{Hypothesis} & \textbf{Support criterion} & \textbf{\shortstack{Reference\\rate}} & \textbf{\shortstack{Power of 1\\comparison at 0.25}} & \textbf{\shortstack{Trials for\\80\% power}} & \textbf{\shortstack{Met by\\chance}} & \textbf{\shortstack{Expected\\count}} \\
\midrule
H1 & At least 3 of 15 (10 vs.\ 60) & 0.95 & 53.2\% & 24 & 4.7\% & 1.5 \\
H2 & At least 3 of 12 (10 vs.\ 10) & 0.50 & 18.2\% & 60 & 1.5\% & 0.5 \\
H3 & At least 2 of 6 (10 vs.\ 60) & 0.95 & 53.2\% & 24 & 3.5\% & 1.1 \\
H4 & At least 2 of 4 (10 vs.\ 60) & 0.95 & 53.2\% & 24 & 1.6\% & 0.5 \\
H5 & At least 3 of 15 (10 vs.\ 60) & 0.95 & 53.2\% & 24 & 4.7\% & 1.5 \\
H6 & At least 2 of 6 (10 vs.\ 10) & 0.50 & 18.2\% & 60 & 4.2\% & 1.3 \\
H7 & At least 2 of 5 (10 vs.\ 10) & 0.50 & 18.2\% & 60 & 3.9\% & 1.3 \\
H8 & At least 2 of 4 (40 vs.\ 60) & 0.95 & 91.4\% & 24 & 1.3\% & 0.1 \\
H9 & 1 pooled (20 vs.\ 20) & 0.50 & 34.1\% & 60 & 4.6\% & 0.4 \\
\bottomrule
\end{tabular}}
\caption{\textit{Power and chance support per hypothesis (Boschloo's exact test, $\alpha=0.05$).}}
\label{tab:power-sensitivity}
\end{table}

Each support criterion requires the smallest number of significant comparisons that keeps its chance probability within a category and configuration, or within a configuration for H8 and H9, at or below 5\% (Table~\ref{tab:power-sensitivity}). We therefore do not adjust individual $p$-values for the number of comparisons. Each hypothesis reports the number of categories and configurations in which its support criterion is met, out of 32 (8 for H8 and H9), alongside the number expected by chance.

\subsubsection{Measures}

We collect the measures in Table~\ref{tab:secondary-measures} for every trial. They support further analysis of agent behavior but have no separate support criteria.

\begin{table}[H]
\centering
{\setlength{\tabcolsep}{4pt}\renewcommand{\arraystretch}{1.1}
\begin{tabular}{l p{0.7\linewidth}}
\toprule
\textbf{Measure} & \textbf{Definition} \\
\midrule
First invocation turn & Turn index of the first skill invocation. \\
\addlinespace
Token consumption & Total tokens consumed in a trial, measured and compared within a task category. \\
\addlinespace
Reasoning tokens & Reasoning tokens the model generated in a trial, as reported by the harness, measured and compared within a task category and configuration. \\
\addlinespace
Turns to completion & Total agent turns in a trial, measured and compared within a task category. \\
\addlinespace
Output-produced flag & Whether the trial produced any task output. \\
\addlinespace
File-read flag & Whether the tool-call trace shows the agent reading a project file before its first skill invocation; for trials without a skill invocation, before its first edit to a project file. \\
\bottomrule
\end{tabular}}
\caption{\textit{Measures collected on every trial.}}
\label{tab:secondary-measures}
\end{table}

\subsubsection{Reporting commitments}

For every comparison specified by a support criterion, we report the difference in rates with its $p$-value and a 95\% confidence interval. Analyses beyond the measures specified in this study will be labeled exploratory. For each conclusion drawn from response text or reasoning summaries, a second author reviews the same trials without seeing the first author's conclusion. Both authors discuss any disagreement and record the final decision. We will report and explain deviations from this registered report.%

\subsection{Data collection}
\label{sec:collection}

Each trial runs unattended through the harness CLI or SDK in a fresh Docker container, with a single task prompt and no working tree, cache, or conversation state retained from earlier trials. The working tree contains only a copy of the selected project directory. Before each trial, we restore the category's untouched snapshot, write the skill pool, and place the artifacts required by the condition.

We collect $k$ trials per experimental condition and 60 per Baseline category in each of the 8 configurations (Table~\ref{tab:configs}). We retain each session that terminates abnormally as a recorded attempt excluded from the analysis, and rerun the trial~\citep{lulla2026buildvsbuy}. We classify each such attempt as an infrastructure failure if it fails before its first model request or ends on a failed model request, and as an agent failure otherwise, and report the number of each per condition. A trial that terminates abnormally on 3 consecutive attempts is reported as missing.

We use the same language runtime throughout collection. Harness settings other than effort remain at their defaults except for specified manipulations. We pin the harness version for the duration of each configuration; if the harness version changes, we repeat that configuration. Within each configuration, we collect the Baseline first, followed by H1 through H9, completing each category before proceeding to the next. Within a hypothesis and category, the trials of all conditions run in an order shuffled with a recorded random seed. All trials for one configuration finish before another begins. Claude Code with the Anthropic model is collected first, because its admission gate fixes the tasks and skills. This order lets later hypotheses use earlier results and allows us to correct detected setup defects before proceeding. If we correct a setup defect, we repeat every trial collected under that setup in the affected configuration.%

Each trial records the repository snapshot hash, task prompt hash, environment-reset confirmation, model version string, harness version string, and, for each model request, the endpoint called and the provider that served it. We report the model and harness versions with each configuration's results. Anthropic states that the weights and configuration associated with a model identifier remain fixed~\citep{anthropic2026modelids}. The OpenAI model has no published dated snapshot; its identifier may refer to different versions during collection~\citep{openai2026sol}. A detected model or harness version change requires repeating that configuration.

We capture tool-call traces as the evidence for confirmatory outcomes, including skill invocations, their turns, and file access. We also retain the agent's response text, including text written between tool calls, to distinguish clarification requests from trials in which the agent does nothing. Anthropic and OpenAI models return only a summary of their reasoning, and only when the request asks for one~\citep{anthropic2026thinking,openai2026reasoning}. With default settings, Codex, OpenCode, and Pi request reasoning summaries from the OpenAI model, and Pi also requests them from the Anthropic model~\citep{openai2026codexmodelsource,opencode2026transformsource,pi2026anthropicsource,pi2026openaisource}. Claude Code in non-interactive mode and OpenCode request none from the Anthropic model, which returns no reasoning text by default~\citep{anthropic2026thinking,opencode2026transformsource}. We record for each trial, including those routed through OpenRouter, whether reasoning summaries were returned. We review the response text and any reasoning summaries to investigate filler invocation and the comparisons selected for exploratory review. Neither determines confirmatory outcomes.

A skill is invoked when a tool call returns its \texttt{SKILL.md} content to the agent or to a subagent it starts. In Claude Code and OpenCode, this is a call to the harness's skill tool naming the skill; in Codex and Pi, it is a successful file read or shell command that names the skill's \texttt{SKILL.md} path and returns at least part of its contents~\citep{anthropic2026agentskills,opencode2026docs,openai2026codexskillprompt,pidev2026}. Either event counts in every harness, including partial reads, and a call that returns an error is not an invocation. We record every distinct skill invoked in a trial and whether the agent or a subagent invoked it. Repeated invocation of a skill counts as 1 distinct entry, and trials with multiple entries retain all of them. Invocation order is not an outcome, tiebreaker, or covariate because it does not establish preference. We log the turn of the first invocation and, for a subagent's invocation, take the agent turn that started the subagent. We report the proportion of trials with multiple distinct invocations and the proportion of invocations made by a subagent in each condition.%

\subsection{Outputs}

On publication, we release the analysis code, per-trial records, authored skill names and descriptions, task texts, and each trial's response text and reasoning summaries. Trial records include outcome coding, repository and prompt hashes, environment-reset confirmation, and model and harness version strings. We deposit these artifacts in a public archive with a persistent identifier and permissive license, provide a copy to reviewers at submission, and cite the archive in the completed manuscript. The shared base repository will be released separately with its protocol.

\section{Estimated cost}

We estimate collection costs from a cost pilot~\citep{lee2026costpilot}. In the first measurement, we sent the prompt "Reply with OK." three times to each configuration in an empty repository and recorded the input tokens of the first API call, which carry the harness's system prompt and tool definitions (Table~\ref{tab:first-call}). Every API call in a session carries this overhead, so session costs include it.

\begin{table}[H]
\centering
{\setlength{\tabcolsep}{5pt}\renewcommand{\arraystretch}{1.1}
\begin{tabular}{l r r r r}
\toprule
\textbf{Model} & \textbf{Claude Code} & \textbf{Codex} & \textbf{OpenCode} & \textbf{Pi} \\
\midrule
Sonnet 5.5 & 19,434 & 15,556 & 7,443 & 2,300 \\
GPT-6.1 Sol & 19,953 & 14,429 & 4,200 & 1,160 \\
\bottomrule
\end{tabular}}
\caption{\textit{Median input tokens of the first API call, by configuration.}}
\label{tab:first-call}
\end{table}

Table~\ref{tab:predicted-cost} gives the estimate from 3 sessions the pilot ran in each of the 8 configurations. In each session, we asked the agent to add request parsing to a working build of the second-stage HTTP server and allowed the session to run until the harness exited. The pilot used the routes in Table~\ref{tab:configs}, except that OpenCode and Pi with Sonnet 5.5 were routed through OpenRouter. We price every API call in each session, including cached input, at the model's standard API prices. Both models have the same standard prices per million tokens, \$2 for input and \$10 for output; we price cached input at \$0.20 per million tokens for Sonnet 5.5 and \$0.10 for GPT-6.1 Sol, and cache writes at \$2.50 for both~\citep{anthropic2026pricing,openai2026pricing}. Each configuration's median and mean session costs give its cost per trial, and its study cost is that cost multiplied by its share of the 20,400 trials: 2,760 for Claude Code with Sonnet 5.5 and 2,520 for each other configuration. Totals are summed before rounding. The repository reports the input, cached-input, and output tokens of every API call, the first-call measurements, and the prices used.%

\begin{table}[H]
\centering
{\setlength{\tabcolsep}{5pt}\renewcommand{\arraystretch}{1.1}
\begin{tabular}{l l r r r r}
\toprule
 & & \multicolumn{2}{c}{\textbf{Session cost (\$)}} & \multicolumn{2}{c}{\textbf{Study cost (\$)}} \\
\cmidrule(lr){3-4}\cmidrule(lr){5-6}
\textbf{Model} & \textbf{Harness} & \textbf{Median} & \textbf{Mean} & \textbf{Median} & \textbf{Mean} \\
\midrule
Sonnet 5.5 & Claude Code & 0.150 & 0.149 & 413 & 410 \\
 & Codex & 0.112 & 0.114 & 282 & 288 \\
 & OpenCode & 0.080 & 0.080 & 202 & 200 \\
 & Pi & 0.054 & 0.092 & 136 & 231 \\
GPT-6.1 Sol & Claude Code & 0.153 & 0.155 & 384 & 390 \\
 & Codex & 0.062 & 0.060 & 156 & 152 \\
 & OpenCode & 0.064 & 0.061 & 161 & 155 \\
 & Pi & 0.055 & 0.055 & 138 & 139 \\
\midrule
\textbf{Total} & & & & \textbf{1,873} & \textbf{1,966} \\
\bottomrule
\end{tabular}}
\caption{\textit{Estimated collection cost by configuration, from the cost pilot's session costs.}}
\label{tab:predicted-cost}
\end{table}

\section{Threats to validity}

\paragraph*{Construct Validity.}
H1--H8 measure invocation outcomes across repeated trials; they do not assess whether an invoked procedure is followed correctly or improves the work. For H1--H8, a trial where the target skill is invoked and the task is completed well scores identically to one where it is invoked and the task is completed poorly. H9 instead measures the effect of skill availability on action occurrence, without isolating whether that effect depends on invocation.

\paragraph*{Internal Validity.}
Adding metadata detail or pool entries increases both information content and text length, which the design does not separate. Padding shorter conditions to equalize length would itself add content that could affect routing.

The target's position in the skill list shown to the agent may also affect invocation. \citet{liu2024lostinthemiddle} found lower retrieval accuracy for entries in the middle of serialized key-value lists than at either end. OpenCode sorts the list by skill name; the cited Codex and Pi source files sort directory entries without establishing the order of the full list~\citep{openai2026codexsource,opencode2026skillsource,pi2026skillsource}, and we have not established Claude Code's ordering rule. The design does not control list position.

Each harness combines a system prompt, tool set, context assembly, and skill presentation. Differences between harnesses may reflect any of these features, which the study does not isolate. The cross-vendor configurations also use an aggregator to translate requests between vendor APIs~\citep{openrouter2026}. Results for these configurations may therefore depend on translation as well as the model and harness.

We collect the 8 configurations sequentially. Anthropic guarantees fixed weights and configuration for a model identifier~\citep{anthropic2026modelids}, but this does not establish that serving infrastructure, routing, or sampling stays unchanged. Without a published dated snapshot for the OpenAI model~\citep{openai2026sol}, comparisons between its harness configurations may also reflect model updates. These risks apply to comparisons across configurations, including counts of configurations supporting a hypothesis. In the cross-vendor configurations, differences between upstream providers selected by the aggregator may also affect results.

Only Claude Code with the Anthropic model must pass the admission gate. We revise tasks and skills until this configuration passes, so their wording may suit it better than the other 7 configurations and raise its invocation rates. This may favor that configuration in comparisons across configurations.

\paragraph*{External Validity.}
All study objects are synthetic. The HTTP server has no third-party dependencies or project history, and its functionality is limited to serving requests. These omissions remove context that may affect routing in a deployed repository. We draft skills using marketplace examples but admit them only when their fields contain the information specified for their conditions. The measured rates therefore apply to this repository, the authored pool, and the single task in each category. They do not estimate invocation rates in established codebases or pools taken from a marketplace.

Models may encounter tool interfaces and skill formats during training. For example, R2E-Gym trained agents on trajectories using a file editor, a search tool, a bash terminal, and a submit action~\citep{jain2025r2egym}. The study's skills are newly authored, but similar metadata, procedures, or formats may occur in the models' training data.

\paragraph*{Conclusion Validity.}
Identical inputs do not guarantee identical outputs from stochastic coding agents. \citet{ouyang2024nondeterminism} found that, for 47.6--75.8\% of tasks, no test input produced equivalent outputs across all 5 programs generated from identical prompts. \citet{bjarnason2026randomness} found variation in multi-step agent runs, with single-run pass rates differing by 2.2--6.0 percentage points depending on which run was reported. Our trials similarly involve multiple agent steps, with sampling parameters such as temperature left at the harness or vendor default, and \citet{blyth2024creativecorrect} showed that temperature alone measurably shifts pass rate and output diversity in code generation. Each additional repetition adds a trial, and its cost, to every condition in every category and configuration. Most conditions therefore have 10 repetitions, which limits the precision of rate estimates; the estimates may differ in an independent replication.

The power estimates in Table~\ref{tab:power-sensitivity} are lowest for the 10-versus-10 comparisons used in H2, H6, and H7. A non-significant result does not establish the absence of an effect. In H9, opposite changes in test execution and docstring updates can offset each other in the pooled comparison, so we report the difference for each action descriptively. Comparisons that narrowly miss significance receive the exploratory review described in Section~\ref{sec:analysis}.

\bibliographystyle{ACM-Reference-Format}
\bibliography{refs}

\end{document}